# Power Allocation in Two-way Relaying Networks

Gang Wang

College of Information Science and Engineering, Changsha University,410000,ChangSha Chnia

**Abstract**.

In this paper, we study relay selection and power allocation in two-way relaying networks consisting of a source, a destination and multiply half-duplex decode-and-forward (DF) relays. A transmission model with three time subslots is purposely introduced. In the first subslot, selected relay applies time-switching protocol to harvest radio frequency energy radiated by source and destination; in the remaining subslots, selected relay facilitates source and destination to exchange information. Due to finite-size data buffer and finite-size battery of relay, an optimal relay selection and power allocation policy is proposed, in order to maximize networks sum-throughput. One obstacle is the inherent non-convex property of the underlying sum-throughput optimization problem. By carefully decoupling the multiplicative variables and relaxing binary variable to a real number, we convert this problem into a convex optimization one and then Karush-Kuhn-Tucker (KKT) conditions are used to solve it. Extensive simulations have been conducted to demonstrate the improved sum-throughput with our proposed strategy.

## 1 Introduction

Relay communication technology is a promising way to improve system capacity, resist channel fading and extend communication coverage [1]. Once one of two links at relay is degraded, the traditional relay transmission mode will not be able to achieve desired performance. For this reason, the so-called buffer-aided network has been proposed in [2]. It is able to store the received data in the buffer, not forwarding the data until the channel state condition is good. In [3], the authors proposed three buffer-aided adaptive relay transmission schemes considering FD relaying networks. In [4], the authors considered buffer-aided successive relaying networks with HD relays. Generally speaking, there are two kinds of forwarding protocols at relay side, namely, amplify-and-forward (AF) and decode-and-forward (DF). In two-way relaying networks, outage probability of AF, DF and hybrid AF-DF relay was analyzed respectively in [5]. However, wireless nodes are powered by limited power beacon. In order to make networks more freedom to choose suitable relay to maximize sum-throughput, and reduce energy burden of a relay, relay selection becomes very important. On the other hand, power allocation can avoid energy waste and allocate appropriate transmission power according to current energy storage. In [6], the authors considered energy-efficient power allocation in the downlink. In [7], the authors proposed an online buffer-energy-aware adaptive scheduling scheme to improve networks throughput.

## 2 Related Work

In the two-way relaying networks, two users exchange information through a relay [8]. Bidirectional relay channel is proposed, enabling source-relay and destination-relay transmission phases into one phase, namely multiple access phase; in broadcast phase, relay-source and relay-destination are also combined into one phase. Optimal power allocation was proposed under the power constraint [11], and relay selection was proposed with energy harvesting relay [9]. In [10], the authors proposed relay selection and power allocation policy to improve secondary networks throughput. There are two transmission modes, including fixed rate transmission [11], and adaptive rate transmission [12]. In [13], the authors considered the silent state of nodes, adaptive mode selection were analyzed where relay has two data buffers. Reference [12] extended this situation to delay-constrained case. In [14], the authors designed an adaptive transmission mode between relay receiving phase and relay transmitting phase. In [15], the authors put forward the energy efficiency model under asymmetric service requirement for two users. Considering delay constraint, transmission mode selection was introduced in [16], relay selection policy was proposed in [17], discrete transmission rate was considered in [18].

In recent years, the combination of buffer-aided communication and relay technology has been widely concerned, buffer-aided relaying protocol enables system to adjust transmission strategy adaptively according to the channel conditions. With the help of data buffer, system adaptively switched between two transmission modes, namely source information transmission mode and relay harvest-and-transmit mode . Similarly, system can adaptively adjust the transmission power to maximize achievable average rate 8. However, buffer-energy-aware adaptive transmission strategy was put forward under energy storage and data buffer constraints. In practice, buffer size is finite. In [19], under the finite-size battery and data buffer, optimal throughput of offline and online was proposed. The effect of battery and data buffer size on system throughput was considered in [20]. When data buffer is involved, delay tolerant and delay constraint are considered. Considering delay constraint protocol, the achieved delay and throughput was studied by modeling the Markov chain queues on the buffer.

## 3 Networks Model

Cooperative communication protocols can be divided into fixed relay strategy and adaptive relay strategy. In the fixed relay strategy, the channel resources from the source node to the relay are divided in a fixed way. Adaptive relay strategy considers selective DF relay and incremental relay. The advantages of fixed relay are easy implementation and simple network structure, while the disadvantages are low spectrum efficiency.

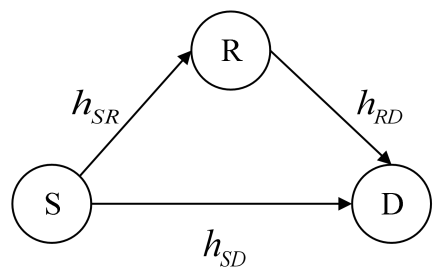

Figure1 Single relay collaboration model

In order to reduce the complexity, we consider the fixed relay scheme with single relay, mainly analyzing the cooperative schemes of amplification and forwarding (AF), decoding and forwarding (DF) and coding and cooperative forwarding (CC). A relay strategy can be described by two orthogonal stages to avoid the interference between the two stages. We consider a three-node network, namely, the source node, the destination node and the relay node. In the first stage, when there is a direct link, the source node broadcasts information to the destination node and the relay; In the absence of a direct link, the information sent by the source node can only be received by the relay; In the second stage, the relay selects the appropriate cooperation mode to forward the information to the destination node, and the destination node decodes it according to the information received in the two stages.

Assume that the signal sent by the excitation source S is, the transmission power is,, and the channel coefficients of the source S to the destination D, the source S to the relay R, and the relay R to the destination D, respectively; And white Gaussian noise, obey the distribution.

## 4 Optimal Relay Selection

Amplification Protocol (AF) was first proposed by Laneman. In this protocol, the relay amplifies the signal (useful signal and noise) received from the source node, the amplification factor depends on the channel state information, and then the relay forwards the amplified signal to the destination node. When there is a direct link, the destination node can receive the information transmitted from the direct link and the copy of the information transmitted from the relay link. The destination node combines these two signals to obtain diversity gain and realize the reconstruction of the original signal.

$$C_1 = \frac{1}{2}\min\{E[\log_2(1+r_{SR}(n)), \log_2(1+r_{RD}(n))]\} \qquad (1)$$

$$C_{SR}(n) = \log_2(1+r_{SR}(n)) \qquad (2)$$

$$Q(n) = Q(n-1) + C_{SR}(n) \qquad (3)$$

$$C_{RD}(n) = \log_2(1+r_{RD}(n)) \qquad (4)$$

$$\min\{C_{RD}(n), Q(n-1)\} \qquad (5)$$

$$Q(n) = Q(n-1) - C_{RD}(n) \qquad (6)$$

The information of the relay source node is directly amplified and forwarded to the destination node, and the destination node performs ideal equalization on the channel fading from the source to the relay. Where is the channel coefficient relayed to the destination node; To relay the information received in the first stage; Gaussian white noise; The amplification factor is inversely proportional to the received power. Wherein, the transmission power of the relay is the noise power.

Under the AF model, information transmission no longer depends on the capacity of the S-R channel, but on the overall capacity. It is suitable for the poor or unstable quality of the S-R channel. The algorithm is not complicated, but it will amplify irrelevant signals.

The forwarding protocol (DF) was first proposed by Sendonaris. The relay decodes and re-encodes the received signal and sends it to the destination node. The DF mode must meet two conditions: one relay node can accurately receive the transmitted signal from the source node, decode it and re-encode it for forwarding; The second destination node receives the transmitted signals of the relay and the source node, combines them in diversity, and can decode them accurately. Only when these two necessary conditions are met can the transmission be successfully completed.

The destination node will combine and decode the signal sent by the source node in the first stage and the signal sent by the relay node in the second stage after receiving it. The maximum ratio combining (MRC) method can be adopted, so the destination node can distinguish the two signals in time and realize the time diversity gain.

The total channel capacity of the system is affected by the channel capacity of S-R and R-D.. If the S-R channel quality is not good, then the relay node can't decode the information transmitted by the source node, so the relay can't play a cooperative role, so DF is suitable for the network with better S-R channel quality. Although DF relay can avoid noise amplification, it may erroneously decode the information of the source node and send it to the destination node, resulting in system performance degradation.

Cooperative coding (CC) protocol was first proposed by Hunter et al. Because DF mode uses simple repetitive codes, the coding efficiency is not high, and the proposal of coding cooperation can bring more coding gains.

Cooperative coding combines channel coding with multiuser cooperative diversity. It divides the codeword of the source node into two parts, which are transmitted by two independent channels (source node and relay node), so the relay can not only decode correctly, but also need to know the coding structure of the source node. The information sent by the relay node will be the information of the source node and additional redundant information. If the relay node can correctly decode the first part of information of the source node, it will send the second part of information of the source node; Otherwise, the relay node will send the second part of its own codeword, that is, the non-cooperative mode.

According to the transmission quantity and direction of information flow on the link, relay cooperation can be divided into one-way relay cooperation and two-way relay cooperation. When both users need to exchange information, only one channel is working in one time slot, while other channels are idle, which will waste spectrum resources. Two-way relay cooperation allows two information flows with opposite directions to be transmitted on one channel, so that the two-way transmission can be completed in two or three time slots, and the spectrum utilization rate is improved. document. The transmission rates of bidirectional relay and unidirectional relay are analyzed. The simulation results show that the spectrum utilization rate and transmission rate of bidirectional relay network are twice that of unidirectional relay wireless network. Therefore, in this paper, we mainly analyze the bidirectional relay network.

Two-way relaying networks (TWRN) models include TWRN with four slots, TWRN with three slots and TWRN with two slots. This part will analyze these three models.

One-way relay network can be expressed as TWRN with four time slots. When there is a direct link between the source node and the destination node, the cooperation between the source node and the relay can form a virtual antenna array to realize diversity multiplexing. The source and destination nodes of unidirectional relay networks can be used as transmitters and receivers, and can be divided into unidirectional relay networks with direct links and unidirectional relay networks without direct links.

A unidirectional relay network can be simplified into a source node S, a relay node R, and a destination node D. Since S-D has no direct link, the source node and relay node will not receive the information from both sides. In the first time slot, the source node sends information to the relay node; In the second time slot, the relay node selects an appropriate forwarding protocol and forwards the information received in the first time slot to the destination node; After these two time slots are completed, the destination node receives the information sent by the source node. The working principle of the third and fourth time slots is the same as that of the first two time slots, and the source node will receive the information sent by the destination node. It takes four time slots for the source and destination nodes to complete the information exchange. When there are multiple source and destination nodes, this working model will cause serious waste of resources.

## 5 Simulation Results

In this section, we analyze the performance of proposed buffer-aided two-way relaying system. We consider a block fading single user Rayleigh fading channel with zero mean and unit variance, noise power is $0.1\ W$. We set transmission deadline $N=5\ s$. The relay R has two finite-size data buffer and an finite-size battery. The networks has 5 relays, i.e., $K=5$. Energy conversion efficiency is 0.8, i.e., ${\eta}=0.8$.

Fig.2 plots the system sum-throughput against the convergence with ${\delta}=10^{-4}$ under different buffer size $B_{max}$ and battery capacity $E_{max}$. Because Lagrange multiplier algorithm is to obtain the optimal solutions by iterating, sum-throughput increases with the number of iterations, and finally converges at 16 iterations. It is observed that sum-throughput increases with the increase of $B_{max}$ and $E_{max}$.

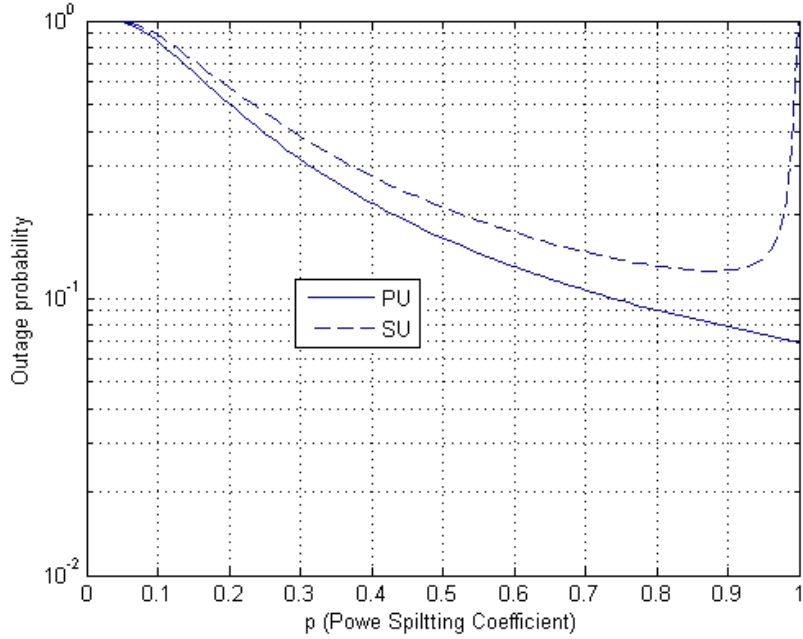

Figure2   Example of embedding vector weighting

Two-way relay network include TWRN with three time slots and TWRN with two time slot. Compared with unidirectional relay wireless network, bidirectional relay network saves at least one or two time slots, because TWRN has joined broadcast and multiple access protocols.

In the traditional wireless communication network, nodes only store and forward. In multicast network, nodes can process the received information and then forward it. The network coding enables the relay node to encode the received data packets and then forward them, and the destination node decodes the complete information of the source node by using certain processing methods. Combining bidirectional relay network with network coding improves the transmission rate of information and solves the problem of spectrum resources.

## 6 Conclusion

In this paper, we proposed relay selection. the source node and the destination node respectively send information to the relay; In the third time slot, the relay network encodes the information received from the source and destination nodes, and then forwards it to the source and relay in the form of broadcast. The source and destination nodes adopt self-interference cancellation to propose useful information. In the first time slot, the source and destination nodes simultaneously send information to the relay in a multiple access mode, and the relay processes the two information; In the second time slot, the relay encodes the forwarded information and broadcasts it.

The source and destination nodes use self-interference cancellation to extract useful information. Resource allocation and outage probability in cognitive two-way relay networks are another research points.